\documentclass[12pt]{article}

\usepackage{newtxtext,newtxmath}
\usepackage{graphicx}

\usepackage[letterpaper,margin=1in]{geometry}

\renewenvironment{abstract}
	{\quotation}
	{\endquotation}

\date{}

\makeatletter
\renewcommand{\fnum@figure}{\textbf{Figure \thefigure}}
\renewcommand{\fnum@table}{\textbf{Table \thetable}}
\makeatother

\usepackage{scicite}

\usepackage{url}

\def\scititle{
    	Distinguishing thyroid nodule malignancy using digital palpation
    }
    \title{\bfseries \boldmath \scititle}

\author{
	Brendan~M.~Unikewicz$^{1}$,
	Joseph~E.~Bonavia$^{1}$,
	Ekaterina~L.~Koelliker$^{2}$,\and
    Abigail~Klein$^{3}$,
	Katie~C.~Spaeth$^{1}$,
	Hannah~M.~Varner$^{1}$, \and
    S.~Kiana~Naghibzadeh$^{4}$,
    Yoav~Horev$^{5}$,
	Sareh~Parangi$^{2\dagger}$, 
    Tal~Cohen$^{1,4\dagger}$\and
	\small$^{1}$Department of Mechanical Engineering, MIT, Cambridge, MA, USA.\and\
	\small$^{2}$Department of Surgery, Massachusetts General Hospital and Harvard Medical School, Boston, MA, USA.\and\
    \small$^{3}$Department of Electrical Engineering and Computer Science, MIT, Cambridge, MA, USA.\and\
    \small$^{4}$Department of Civil and Environmental Engineering, MIT, Cambridge, MA, USA.\and\
    \small$^{5}$Division of Otolaryngology, Head and Neck Surgery, University of Alberta, Edmonton, Alberta, Canada.\and\
	\small$^\dagger$Corresponding author. Email: talco@mit.edu and sparangi@mgb.org\and
}    
    
\begin{document} 

% Insert the title and author list
\maketitle

\clearpage

% Abstract, in bold
% There are strict length limits, and not all formats have abstracts.
% Consult the journal instructions to authors for details.
% Do not cite any references in the abstract.
\begin{abstract} \bfseries \boldmath
 Accurate pre-operative diagnosis of thyroid nodules remains a major clinical challenge. Current diagnostic techniques rely on fine-needle aspiration and demonstrate high rates of inconclusive results, often leading to avoidable thyroidectomies with lifelong implications. Here, we introduce a mechanics-based diagnostic method that leverages Digital Palpation to determine malignancy and identify tissue subtypes in human thyroid nodules. From a cohort of patients undergoing thyroidectomies, we evaluated \textit{ex vivo} thyroid specimens and processed the resulting mechanical data from 36 tests in 10 thyroid nodules. Principal Component Analysis reveals that these experimental results are distributed within a three-dimensional space of interpretable scalar features reflecting tissue \textit{grittiness}, \textit{strength}, and \textit{toughness} which map closely to descriptions of malignant and benign nodules given by clinicians. Our results further demonstrate that malignant nodules exhibit distinct mechanical signatures compared to their benign counterparts. This revealed a statistically significant difference ($p = 9 \times 10^{-4}$) and separability (95\% lower-bound Hedges’ $g = 1.08$) between benign and malignant nodules along a direction in principal component space representing effective \textit{stiffness}. Preliminary clustering patterns were also observed among malignant subtypes, including follicular, oncocytic, and papillary thyroid carcinomas. Together, these results highlight the potential of Digital Palpation as a novel mechanics-based method for tissue characterization and motivate its further development toward diagnostic applications.
\end{abstract}

\clearpage 

\section*{Introduction}
By the age of 60, the majority of adults will present with a thyroid nodule  \cite{li2021prevalence,tamhane2016thyroid,gutman1998fine}, i.e. a discrete lesion within the thyroid
gland that is radiologically distinct from surrounding thyroid tissues  \cite{haugenbryan20162015}. While most nodules ($\sim$90\%) are benign  \cite{durante2018diagnosis}, some may harbor cancer. In 2022, there were an estimated 1,016,930 people living with thyroid cancer (TC) in the United States   \cite{NCISERStatThyroid}, reflecting a significant increase in the incidence of TC, which has tripled in the past three decades  \cite{ctx12198192920006761, yan2020rising, luster2019thyroid, haymart2013variation, seib2019evolving}. 
In the same year, TC was reported as the fifth  most commonly diagnosed cancer worldwide in  women;  particularly affecting young women ($<50$ years), for which it was the second most commonly diagnosed cancer  \cite{ctx12198192920006761, gco-cancer-today-pie-0to17}. 

The increasing prevalence of TC explains an increase in the need for reliable and minimally invasive diagnosis methods for TC. The existing standard of care for TC diagnostics relies on
fine-needle aspiration (FNA) followed by cytology, which has a broad range of reported false negative rates (3-46\%) that increases with nodule size  \cite{vaghaiwalla2025reassessing, nam2017large,rossi2019pitfalls,bongiovanni2012bethesda,richmond2014false,kulstad2016all}, and cannot decisively classify up to one third of nodules  \cite{NatureTCOver, NatureTCOver2, saint2017overdiagnosis, ahn2014korea,davies2014current}, leading patients to undergo molecular testing, and then diagnostic surgery, i.e. a thyroidectomy.   As a result, a large fraction of resected thyroids (40-70\%) ultimately receive a benign pathology result \cite{fna70,MTC_FNA,trimboli2015detection}, implying that  the major surgical intervention may have been avoidable \cite{ho2014malignancy, endo2020features,haugen20162015}. 
Regardless of the final pathology result, the increasing population of patients being treated for TC results in substantial costs to patients, insurers, and society  \cite{mongelli2020financial,lubitz2016changing}. Patients that are treated for TC experience significant effects to their quality of life on par with other cancers  \cite{applewhite2016quality, aschebrook2015risk, mongelli2020financial}. Including the need for life-long hormone replacement and the management of potential surgical complications that may include vocal cord paralysis and hypocalcemia \cite{burge1998risks,liu2017risk,sancho2008risk,christou2013complications,tran2019thyroid}. 

These shortcomings have motivated efforts to improve upon the existing diagnostic methods and the current standard of care. One significant improvement has been achieved by the development of molecular testing (e.g. Afirma and Throseq) \cite{krane2020afirma,jug2018molecular, ohori2019benign, chen2020role, patel2023molecular}, which was shown to reduce the number of diagnostic surgeries in indeterminate nodules \cite{livhits2021effectiveness}; however,  molecular testing relies on successful FNA and remains an expensive and time intensive (2-3 weeks) diagnostic. Notably, some studies report that, in practice,  many patients who receive molecular testing eventually undergo surgical resection  \cite{digennaro2022assessing, wong2020cost, aschebrook2013clinical,steward2019performance, patel2020american, khan2020thyroid}.
Another approach to improve TC diagnosis has leveraged advanced imaging based methods \cite{wells2011medical,nitta2021review,anbarasan2021characterisation} (e.g.\ elastography, PETCT); however, these approaches have not shown improved accuracy over the current standard of care \cite{yoo2018computer, moon2012diagnostic, luster2019thyroid,sigrist2017ultrasound, selvaraj2021diagnostic,anbarasan2021characterisation}, and are thus not currently recommended in the  diagnosis workflow  \cite{patel2020american}. Similarly, the use of Artificial Intelligence (AI) to enhance imaging based disease classification has yet to  offer improved accuracy over the current standard of care  \cite{toro2023artificial,alabrak2023artificial}. 

%To alleviate this issue, there have been several 

%Over-diagnosis  \cite{NatureTCOver, NatureTCOver2, saint2017overdiagnosis, ahn2014korea,davies2014current} and the resulting over-treatment of TC is  a direct result of increased diagnostic scrutiny empowered by the widespread accessibility to diagnostic imaging modalities and FNAC [ \cite{lubitz2016changing, luster2019thyroid, davies2014current}]. 
%Thyroid nodules are detectable in two-thirds of the population [ \cite{tamhane2016thyroid}]. An estimated  300,000 new thyroid nodules are diagnosed each year [ \cite{gharib2004changing}], and at any one time, between 10 and 20 million Americans have clinically detectable thyroid nodules [ \cite{harvey}]. However, of the vast number of detectable thyroid nodules,  only a small fraction  harbor cancer [ \cite{ho2014malignancy, endo2020features}], indolent thyroid cancers or suspected thyroid nodules are often treated as aggressively as symptomatic cancers due to uncertainties associated with current diagnostic tools.
%Recognizing the urgent need to reverse trends in over-treatment, in 2015 the American Thyroid Association for the first time recommended  transition to alternative ‘less is more’ management strategies, including active surveillance for nodules lacking suspicious or aggressive features [ \cite{haugen20162015,NatureTCOver2}]. 
Motivated by the common perception among  clinicians that various malignant and benign nodules are occasionally physically distinguishable to the touch, a nascent and promising approach aims to leverage these reported differences to improve the classification of nodules.  
After handling the tissue during surgery, in the pathology laboratory, or while performing FNA, clinicians often use terms such as  \textit{tough}, \textit{gritty}, \textit{firm}, \textit{hard}, etc. \cite{hoang2015thyroid, azizi2013performance, luo2017risk}, to describe and distinguish nodules. Notably, clinicians often specify that malignant nodules feel particularly \textit{stiff}. Studies have shown that such mechanical indications obtained via manual palpation \cite{rosenzweig2025palpation} add clinical value and can serve as auxiliary diagnostic evidence \cite{hannequin1988multivariate,wu2022clinical,sharma2019piezoelectric, luo2017risk}; however, the highly subjective and qualitative nature of these clinical descriptions limit their diagnostic value. Accordingly, if we could quantify these mechanical descriptions of nodules, diagnosis could improve. 
\begin{figure*}[ht]
    \centering
    \includegraphics[width=\columnwidth]{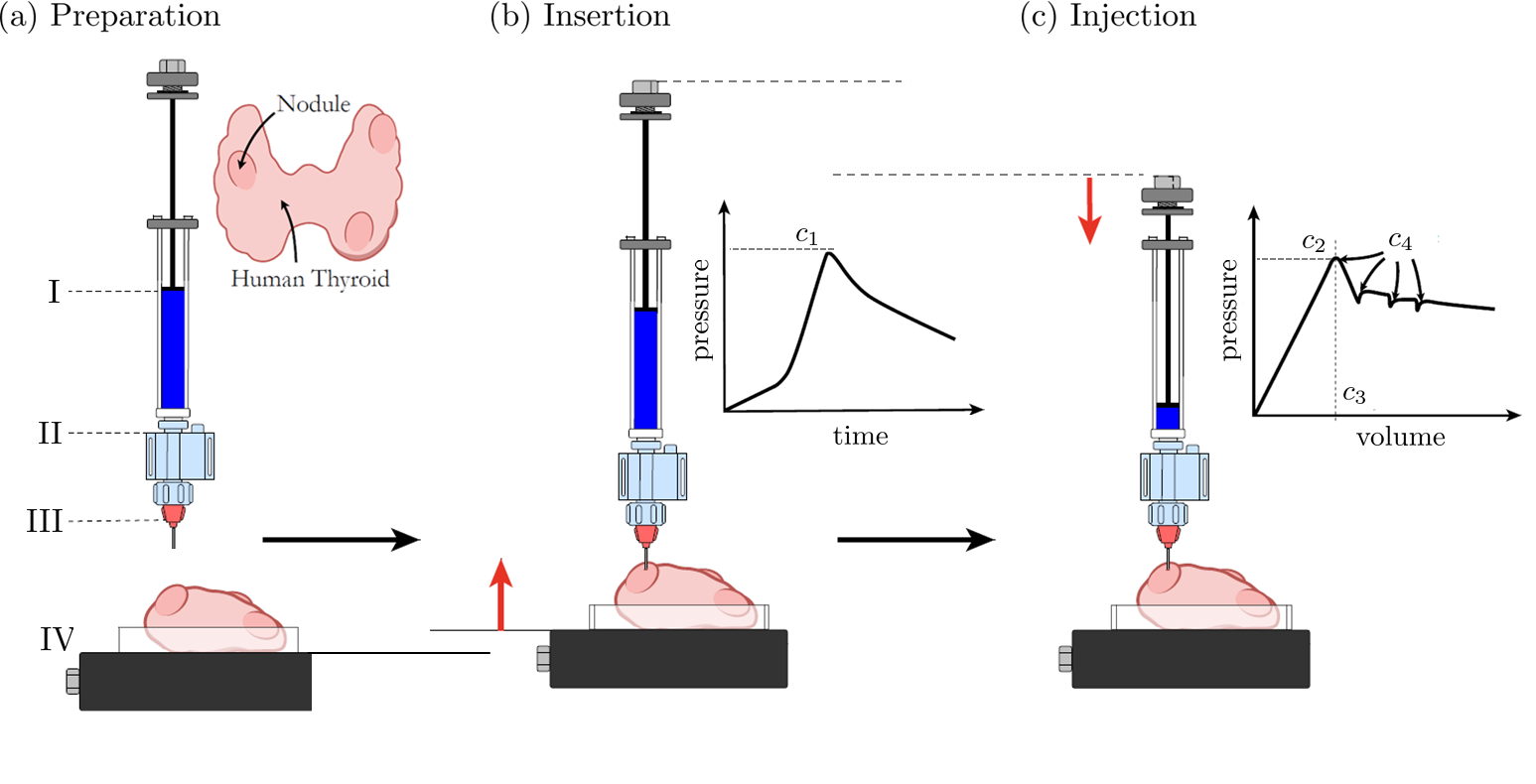}
    \caption{ {\textbf{Schematic overview of the Digital Palpation testing setup and data collection protocol.} (a) Preparation: A fixed syringe assembly (I) is pre-filled with Phosphate-buffered saline and fitted with a pressures sensor (II), and a 25 Gauge hypodermic needle (III).   A resected thyroid specimen containing nodules (inset) is  positioned atop a vertically actuated translation stage (IV) with the interrogated nodule placed directly below the needle. (b) Insertion:     Controlled needle insertion is achieved  by driving the actuated stage  towards the needle. Simultaneous  recording of the pressure produces a pressure-time response referred to as the \textit{Needle Insertion Curve}, as shown schematically (inset). We denote the peak pressure during insertion by $c_1$.  (c) Injection: The fluid injection procedure is subsequently initiated by displacement of the syringe plunger within the syringe assembly, injecting fluid at a controlled rate (with nanoliter precision) into the tissue to expand a millimeter scale cavity. The resistance of the tissue to expansion results in an increase in pressure that is recorded via the pressure sensor to obtain  a pressure-volume response termed the \textit{Cavity Expansion Curve}, as shown schematically (inset). We identify three key characteristic features; the peak   pressure $c_2$, the corresponding volume $c_3$,  and lastly, $c_4$ corresponds to the number of peaks.}}
    \label{fig1}
\end{figure*}

To quantify mechanical heterogeneity in human thyroid tissue lesions Sharma et al.  \cite{sharma2019piezoelectric} employed piezoelectric needle sensors that measure the resisting force upon needle insertion. They reported distinguishable force and heterogeneity values for different tissue types, providing a first indication of mechanically measurable differences; however, the use of force measurements directly behind the needle creates a dependence on the geometry of the nodule and proximity to other tissues (which as such must be known prior to ascertaining a definitive diagnosis in real clinical settings). Ultimately, to be able to provide robust and repeatable results, a mechanical test for distinguishing and diagnosing thyroid cancer must provide a local measure of material response within the nodule that is independent of the macroscopic geometry of the thyroid.

In this study, we show that \textit{Digital Palpation}; a method to quantitatively characterize the local mechanical resistance of a nodule, independent of its macroscopic geometry or physical location; can distinguish freshly excised human thyroid nodules. This offers a new pathway for improved diagnosis and risk stratification of thyroid nodule malignancies. By measuring the resisting pressure developed within the water column in a sharp injection needle, as it is driven through the tissue layers and ultimately into the nodule, the tearing resistance (amongst other properties) of the tissue is captured. Then by expanding a nanoliter-sized fluid bubble with concurrent pressure measurement, the mechanical resistance to cavity expansion is measured. This multivariate mechanical profile from 6 thyroid specimens, obtained from 36 tests in 10 nodules, was examined via principal component analysis (PCA) \cite{greenacre2022principal} and logistic regression. Signed Euclidean distances from the logistic regression decision boundary were then computed to visualize dimensionality reduction and highlight clusters between benign and malignant nodules, as well as among malignancy subtypes including follicular (FTC), oncocytic (OTC), and papillary thyroid carcinoma (PTC). We also note that the signed Euclidian distance measure separating malignant and benign nodules aligns with clinical descriptions of \emph{stiffer} malignant nodules from manual palpation without assuming so \textit{a priori}.

\section*{Results and Discussion}

\begin{figure*}[!ht]
    \centering
    \includegraphics[width=\columnwidth]{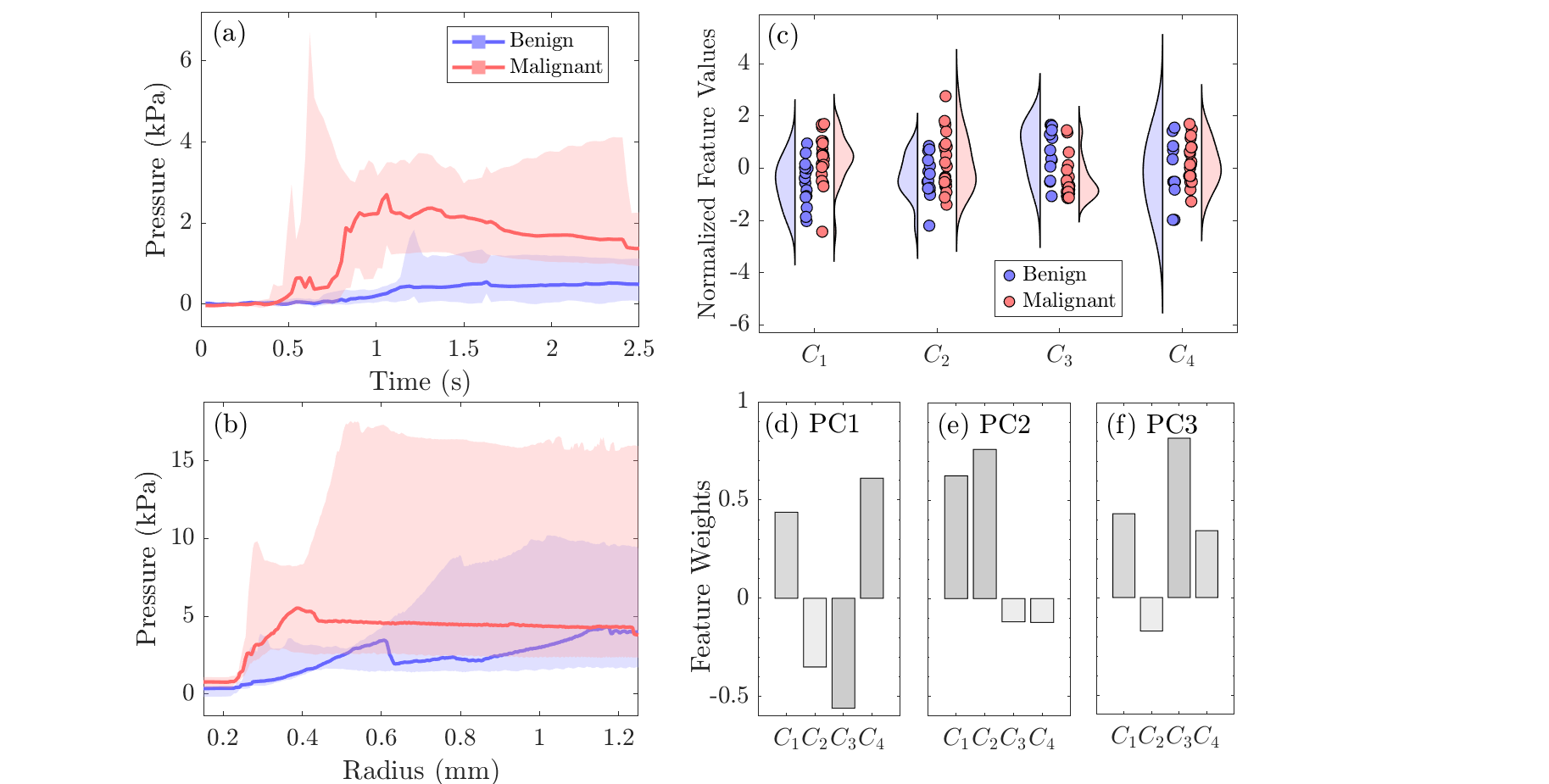}%{figures/FIG2-PNG.png}
    \caption{{\textbf{Raw DP data processed to obtain normalized features and   principal components.}} (a) \textit{Needle insertion} data, and (b) \textit{cavity expansion} data, differentiated between benign (blue) and malignant (red) nodules. In both graphs, the solid lines correspond to the median-averaged value of each group and the shaded regions correspond to the inter-quartile range, respectively, to give an idea of variability between tests. (c) Extracted features $c_i$  are first  log-transformed $(C_i=\ln (c_i))$ and then z-score normalized, such that each feature distribution for benign (red) and malignant (blue) nodules has zero mean and unit standard deviation. (d-f) Principal component analysis results. Feature loadings for the first three principal components illustrate the relative contribution of each normalized feature $(C_i)$. These components may be interpreted as effective measures of  \textit{grittiness} (PC1), \textit{strength} (PC2), and \textit{toughness} (PC3).}
    \label{fig2}
\end{figure*}

\subsection*{Digital Palpation (DP) enables collection of mechanical data in freshly excised human thyroids} The reports from physicians mentioned above include various terminologies to describe complex differences between nodule types obtained  via \textit{manual palpation} or  during FNA testing. Although these reports motivate our effort to quantitatively distinguish nodules via their mechanical response, these reported differences cannot be accurately described using conventional mechanical measures, such as stiffness or toughness alone. Hence, in this work, we depart from common experimental approaches to  study the mechanical response of  biological tissue, which is usually geared towards the development of constitutive models. Instead,  we aim to utilize the mechanical data to maximally distinguish and classify tissue types in a rapid and minimally invasive way. To this end,  we developed a method to quantitatively interrogate the local mechanical resistance of the tissue, at the tip of a needle, throughout both needle insertion and cavity expansion.  

Our approach, which we refer to as \textit{Digital Palpation} (DP), builds on and combines two recently developed methods: needle-based insertion \cite{naghibzadeh2026novel}, and volume-controlled cavity expansion \cite{raayai2019volume,chockalingam2021probing,unikewicz2025accessible} (VCCE); both of which are designed to obtain objective mechanical data that is independent of the macroscopic geometry. Notably, needle-based insertion has been shown to reliably distinguish between soft synthetic materials \cite{naghibzadeh2026novel} and to inform best needle insertion protocols, and VCCE has been successfully applied to characterize not only synthetic materials, but also a variety of biological materials; showing a robust ability to distinguish gray matter from white matter brain tissue \cite{mijailovic2021localized}, changes in blood clots coagulated on different days \cite{varner2023elasticity}, and aging of liver  \cite{varner2025evolving}, highlighting their potential for use in clinical settings. 

As explained in greater detail in the \textit{Materials and Methods} section, surgically excised thyroid specimens exhibiting thyroid nodules were evaluated at Mass General Brigham Newton-Wellesley Hospital using DP with IRB approval from the Mass General Brigham IRB. As seen in Fig. \ref{fig1}a, freshly resected thyroid was placed on the DP translation stage, and positioned directly underneath the needle-plunger system.  Controlled needle insertion was achieved by raising the  translation stage (Fig. \ref{fig1}b)  at a constant rate of 12.50mm/s, targeting the center of the thyroid nodule of interest. The target insertion depth was informed by the overseeing clinician, taking into account the nodule’s distance from the thyroid surface and whether the specimen was repositioned or reoriented to avoid repeated insertions at similar locations (see Tables S1 and S2 in the Supplementary). During insertion, the variation of pressure  within the fluid column  was recorded to produce the \textit{insertion curve} as seen in the inset of Fig. \ref{fig1}(b). Once insertion was complete, the injection of the phosphate buffered saline  (PBS) water solution began at a volumetric flow rate of 0.6$\mu$L/s until a total fluid volume of 9.2$\mu$L was transferred. This fluid volume is equivalent to a spherical cavity with final radius of 1.30mm.

\subsection*{Characteristic Features Emerge in DP Data}\label{statistics}
The data collected via the DP protocol described above may be seen in the Supplementary. From these datasets, we observe typical features that emerge in both the \textit{needle insertion} and \textit{cavity expansion} curves, as illustrated schematically in Fig. \ref{fig1}. For the present analysis, we selected four key features $\{c_1,c_2,c_3,c_4\}$. From the needle insertion curve, we extracted the peak pressure $c_1$ achieved over the entire insertion process, which may be thought of as the maximal resistance of the nodule to needle penetration. From the cavity expansion curves we selected the remaining three features. In a typical cavity expansion process, the pressure increases monotonically as fluid volume increases up until a peak value. Similar to the needle insertion curve, we extract the peak pressure  $c_2$ during fluid injection and the corresponding injected fluid volume at this point $c_3$. These features  represent the transition between elastic expansion to localized tissue rupture. Lastly, the number of discrete peaks that in the injection data, $N$, is also recorded, which may be thought of as a measure of tissue granularity (or heterogeneity), i.e. the number of rupture events as the fluid cavity expands. Due to $N$'s discrete nature and the existence of true zeroes, this feature is  represented by the regularized quantity $c_4=N+\sqrt{1+N^2}$. In what follows, we will show that these four scalar features $\{c_1,c_2,c_3,c_4\}$ provide sufficient  mechanical data to classify thyroid  nodules.  Further details on the extraction of these features from the needle insertion and cavity expansion curves may be found in the \textit{Materials and Methods} section of the Supplementary. Importantly, no leaks or nodule targeting issues were identified in any of the tests. Hence, none of the tests met criteria for exclusion and all data recorded with the aforementioned needle insertion and cavity expansion protocol were used in the analysis.

In Fig. \ref{fig2}(a,b) we examine the differences between data that corresponds to nodules that later received a malignant pathology result (red), from data of those that were later found to be benign (blue). For both needle insertion and cavity expansion the corresponding averaged curves are shown via solid lines and the shaded regions represent inter-quartile range. While   qualitative  differences between groups are visually apparent, substantial overlap does exist, and the variability in each dataset is large. These differences thus motivate the use of statistical methods to better quantify and classify these results. %We thus rely the need for dimensionality reduction through extracted scalar features to better elucidate the underlying separability between tissue types.   

\subsection*{Analyzing DP Data Via Principal Component Analysis (PCA)}
PCA was chosen due to its growing use within the field of classification \cite{greenacre2022principal}. To conduct the PCA we performed a two step normalization procedure. As $\{c_1,c_2,c_3,c_4\}$ were observed to be log-normally distributed, we first switched to the logarithmic values $C_i=\ln(c_i)$. Note that this implies $C_4=\mathrm{asinh}(N)$, a form often used to Gaussianize skewed, heavy-tailed data, particularly when zero-crossings are present. Next, each of the logarithmic $C_i$ values was z-score normalized, as shown in Fig. \ref{fig2}(c), which also reveals observable differences between the coordinate values for malignant and benign data. 
Prior to performing PCA, we confirmed the absence of multicollinearity  \cite{kim2019multicollinearity} and thus the stability of the covariance structure necessary for the validity of this analysis (see \textit{Materials and Methods}).

With the normalized data at hand, PCA was  used to identify an alternative set of values which, through linear combinations of $C_i$, correspond to maximal variance in the data. For our PCA we utilize data from all 36 tests independently, as the goal is to capture the space in which tests vary.  We find that the resulting first three \textit{principal components} (PCs) (Fig. \ref{fig2}(d-e)) capture a majority (88.1\%) of the variance in the dataset (PC1 [45.3\%], PC2 [27.3\%], PC3 [15.5\%]). 

Examining the weights of the different features for each principal component in Fig. \ref{fig2}(d-f) highlights major contributions and allows us to draw additional physical insights into thyroid nodule mechanics. Taking advantage of the logarithmic  normalization we can accordingly write the following power law relationships
% \begin{equation} 
% pc1= \frac{c_1^{0.4}c_4^{0.6}}{c_2^{0.3}c_3^{0.6}},~~~{pc2} = \frac{c_1^{0.6}c_2^{0.8}}{c_3^{0.1}c_4^{0.1}},~~~{pc3} = \frac{c_1^{0.4} c_3^{0.8}c_4^{0.3}}{c_2^{0.2}}
% \end{equation} 
\begin{equation}
\begin{aligned} 
pc1=&\, \frac{c_1^{0.4}c_4^{0.6}}{c_2^{0.3}c_3^{0.6}}\qquad \to \quad \text{effective \textit{grittiness}},\\ {pc2} =&\, \frac{c_1^{0.6}c_2^{0.8}}{c_3^{0.1}c_4^{0.1}}\qquad \to \quad \text{effective \textit{strength}},\\{pc3} =&\,\frac{c_1^{0.4} c_3^{0.8}c_4^{0.3}}{c_2^{0.2}}\ ~ \to \quad \text{effective \textit{toughness}},
\end{aligned}
\end{equation}
where we have rounded the coefficients for compactness and have defined PC$i=\ln({\rm pc}i)$ (see Table S3 in Supplementary). 

Accordingly, pc1 contrasts  contributions from critical insertion pressure and number of discrete rupture events against  contributions from critical expansion pressure and critical expansion volume, with the number of discrete rupture events having the most significant role. Hence, we  interpret pc1 as an effective measure of material \textit{grittiness}. This aligns with the results of Sharma et al.  \cite{sharma2019piezoelectric}, who have shown that ``heterogeneity'' (as determined by the number and size of rupture events) was a significant indicator for understanding different tissue types. 

The second principal component, pc2, increases with the critical expansion and critical insertion pressures, and contains less dominant contributions from the critical expansion volume and the number of discrete rupture events. Since this component highlights the contributions from critical pressures we interpret pc2 as an effective measure of material \textit{strength}. 

Finally, the third principal component, pc3, increases with critical expansion volume, critical insertion pressure, and discrete rupture events, with a small contrasting contribution from the critical expansion pressure. We thus interpret this component as an effective measure of material \textit{toughness}, reflecting the tissue's capacity to accommodate deformation and dissipate energy before failure. The interpretations of the effective \textit{grittiness} (pc1), \textit{strength} (pc2), and \textit{toughness} (pc3), provide physically intuitive axes along which to examine the mechanical distinctions between benign and malignant thyroid tissues, as well as among malignancy subtypes. 

This is encouraging as it seems that quantitative measures analogous to the material properties which surgeons noted varied qualitatively across thyroid nodules capture the majority of variance in our data. Note here that a material \textit{stiffness}, which surgeons often expressed  as one of the most obvious factors between malignant and benign tumors, has  yet to emerge. However, so far all analysis has been agnostic to the malignancy (or lack-thereof) of the nodules. In the following sections we will see that such a measure of \textit{stiffness} does emerge as a deciding factor in nodule malignancy. %Importantly, given the distribution of variance among these three components, it becomes apparent that no single mechanical characteristic can effectively distinguish between nodules, as we show next by visualization of the data. 

\begin{figure}[!ht]
    \centering
    \includegraphics[width=0.5\columnwidth]{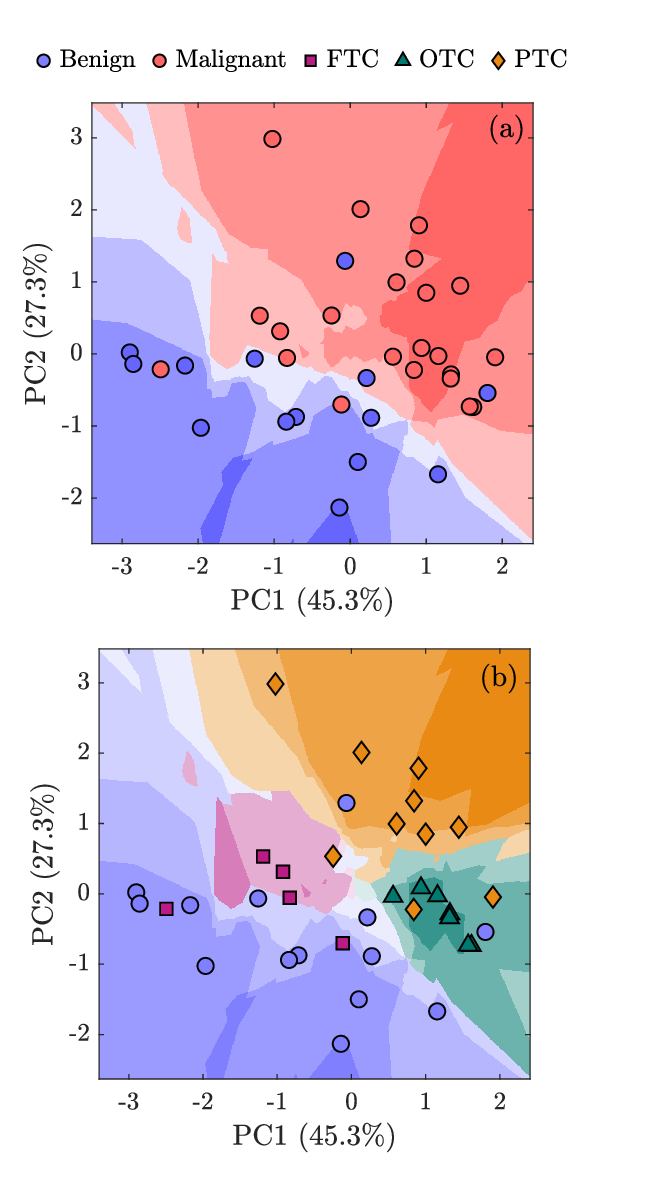}
    \caption{\textbf{Visualization of separation between benign and malignant subtypes.} Data from the PCA are displayed in the PC1-PC2 plane using the k-Nearest Neighbors (k-NN) algorithm with $k=6$ to color regions. (a)  Regions that correspond to benign  and malignant  pathology results are visually separated. (b) Malignancy subtypes are also seen to form clusters. }
   \label{fig3}
\end{figure}
\begin{figure*}[!ht]
    \centering
    \includegraphics[width=\columnwidth]{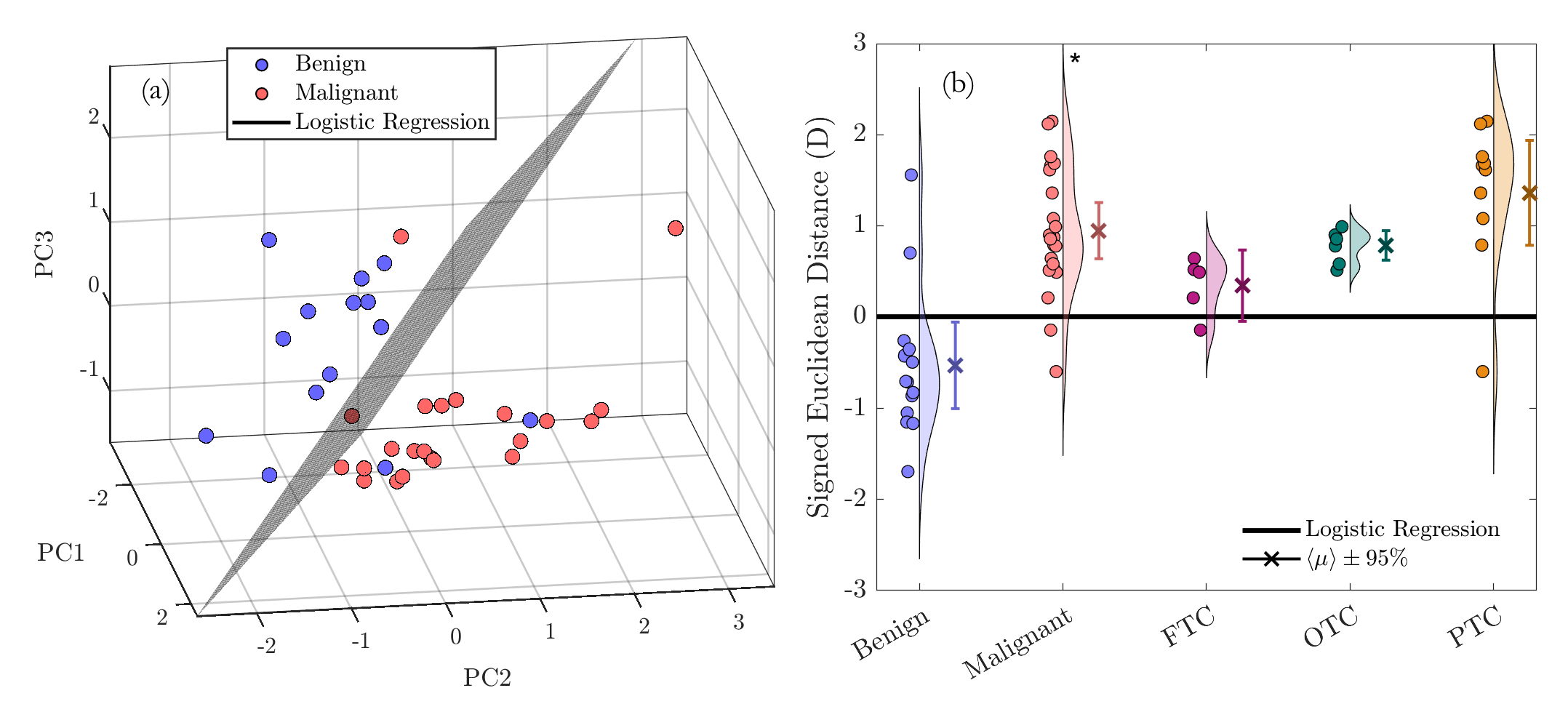}
    \caption{\textbf{Visualization of PCA, logistic regression, and signed euclidean distance differentiation between benign and malignant thyroid nodules.} (a) Data from the PCA are displayed for the first three principal components, capturing 88.1\% of the total variance. A solid grey plane denotes the logistic regression decision boundary separating benign (blue) and malignant (red) thyroid nodules. From the 3D PCA, we can directly translate each plot into a 1D representation using the signed Euclidean distance. (b) The 1D representation highlights that the benign versus malignant separation shows clear distinguishment with their mean $\langle \mu \rangle$ and 95\% confidence interval. Additionally, using the same logistic regression applied for the benign versus malignant separation, we see distinguishment between benign and malignancy subtypes as well.}
    \label{fig4}
\end{figure*}

\subsection*{Nodule Subtype Data is Visually Distinguishable}
Order reduction (i.e from four features for each DP test to three principal components) enables a more intuitive visualization of the data, which can aid in the identification of mechanically based separation between nodules that were later found to have different pathologies (see Tables S1 and S2 in the Supplementary). In both graphs in Fig. \ref{fig3}, each marker corresponds to one test and the marker shape represents the final pathology result: Papillary Thyroid Carcinoma (PTC), Oncocytic Thyroid Carcinoma (OTC), or Follicular Thyroid Carcinoma (FTC). In Fig. \ref{fig3}(a), we consider two sub-groups, coloring benign pathologies in blue and malignant pathologies in red. To further assess whether separation exists, a $k$-nearest neighbors ($k$-NN) analysis with $k = 6$ was applied to the 3D data and visualized as a projection in the PC1--PC2 plane. This visualization revealed a clear separation between malignant and benign nodules.

To explore whether DP can differentiate between malignant subtypes themselves, Fig.~\ref{fig3}(b) presents the same $k$-NN projection with each marker color corresponding to a distinct malignancy subtype (PTC, OTC, FTC). Natural clustering of malignancy subtypes is already observable in this visualization, suggesting that Digital Palpation has strong potential not only to distinguish malignant from benign tissue but also to discern between distinct malignancy subtypes in principal component space. Given our limited sample size, we avoid pairwise regressions between malignancy subtypes to prevent overfitting. Instead, this $k$-NN visualization serves a critical illustrative purpose: clustering subtypes based on their mechanical properties is possible and justifies future work focused on increasing available mechanical data from thyroid nodules, which could further improve differentiation between subtypes

\subsection*{Statistical analysis of Malignancy Separation}
A logistic regression model was chosen as the simplest tool to highlight the binary linear classification of whether a nodule was malignant or benign (Fig. \ref{fig3}(a)). To perform this linear regression we again utilize the data from all 36 tests to capture the boundary between malignant and benign data. Turning to a 3D representation of the DP data in Fig. \ref{fig4}(a), we find that the grey plane, obtained via logistic regression, delineates the statistical boundary between these two groups. 
To quantify this separation, we computed the signed Euclidean distance from each test point to the logistic regression decision boundary plane. This metric provides a 1D measure of both significance ($p$-value) and separability (Hedges' $g$) where positive distances lie on the malignant side of the boundary and negative distances on the benign side (see \textit{Materials and Methods} for more information). As seen in Fig. \ref{fig4}(b) the comparison between benign and malignant data using this approach yielded statistically significant separation with a $p$-value of $p = 1 \times 10^{-6}$ and $g = 1.94$, confirming that the logistic regression designed to separate these two groups retains strong discriminatory power. 

Additionally, we assess averaged nodule-level values to verify that the observed trends are not an artifact of intra-nodule dependence among repeated measurements. For this, we average the signed Euclidean distance of all 10 benign and malignant nodule values. Then we compare the five averaged benign nodules against the five averaged malignant nodules (visualization of the nodule-level data may be seen in Fig. \ref{fig:fig2-esi} in Supplementary). A two sample t-test on nodule-level mean signed distances yielded $p = 9 \times 10^{-4}$ with a Hedges' $g = 2.90$ and 95\% confidence bounds of $[1.08,\ 4.66]$ calculated using a non-central t test. Note that even the lower bound of this Hedges' $g$ estimate is greater than 0.8. This $p$-value indicates that the separation between benign and malignant groups is statistically significant, while the Hedges' $g$ indicates that the magnitude of this separation is substantial.

To account for multiple measurements being collected from the same nodule, we used a permutation-based statistical test that treated each nodule as a single independent unit of analysis \cite{lehmann2005testing}. This test asks, ``If the true diagnostic label were unrelated to the measured features, how often would a random reassignment of nodules to diagnostic categories produce a separation as strong as our original assignment?'' The permutation test yields a $p$-value of $p = 2/252=8 \times 10^{-3} $. This is statistically significant and represents the strongest possible evidence achievable given our sample of 10 nodules. This confirms that our findings are robust to the nested structure of the data (multiple samples per nodule) and that the observed separation reflects genuine biological differences between benign and malignant nodule types, rather than chance clustering of individual measurements.

\subsection*{A Stiffness Metric Emerges from the Multivariate Analysis}

The logistic regression decision boundary separating benign from malignant nodules defines a plane in the space of the first three principal components. Specifically, the signed Euclidean distance ($D$) from any point to this plane in log space provides a scalar measure of malignancy risk, with positive values lying on the malignant side of the boundary. Because each principal component $PCi$ is itself a linear combination of the original normalized features $C_i$, this distance can be expressed directly in terms of the measured mechanical quantities as 
\begin{equation}
d = \frac{c_1^{0.4} \, c_2^{0.6}}{c_3^{0.7}} \to \quad \text{effective \textit{stiffness}},
\end{equation} 
where we have rounded the coefficients for compactness and defined $D = \ln(d)$. Note that the weight of $C_4$ is much smaller than that those of $\{C_1,C_2,C_3\}$ and was thus rounded to 0. This expression reveals that the primary axis of discrimination in our data is an effective measure of material \textit{stiffness}: a ratio of pressures (needle insertion and cavity expansion) to the corresponding rupture volume during cavity expansion. Samples with higher values of this metric, i.e. those that develop higher pressures relative to the volume at which failure occurs, are more likely to be malignant. Remarkably, this mechanically-intuitive quantity emerges naturally from the multivariate analysis, demonstrating that the underlying physical difference between benign and malignant nodules can be captured by a single scalar reflecting the tissue's effective stiffness and directly aligning with clinical intuition where clinicians often report malignant nodules as \textit{stiffer} than benign nodules. This finding not only validates the mechanical basis of our approach but also provides a simple, interpretable metric with clear translational potential.

\section*{Conclusion}
In this proof of principle work we show that mechanical data obtained via digital palpation (DP) exhibits strong potential as a minimally invasive method to inform thyroid cancer diagnosis. A straightforward, low-order statistical analysis of  data from 36 \textit{ex vivo} tests in 10 nodules (from 6 resected human thyroids), reveals clear statistical separation between benign and malignant nodules (p $\le$ 0.05) with a large effect size (Hedges’  g $\ge$ 0.8). Interestingly, this analysis yields a three-dimensional space of \textit{interpretable} quantities that map closely to commonly reported clinical descriptions (i.e., \textit{grittiness}, \textit{strength}, and \textit{toughness}). Performing logistic regression between benign and malignant nodules in this 3D space, we find that differentiation occurs along a single \textit{interpretable} scalar parameter representative of \textit{stiffness}.

Beyond the benign--malignant distinction, preliminary clustering (via k-nearest neighbors) in PC space was observed for the malignancy subtypes: follicular, oncocytic, and papillary thyroid carcinoma. These findings motivate the further development of DP not only as a method for distinguishing benign from malignant thyroid nodules, but also as a potential approach for identifying malignancy subtypes for improved clinical treatment strategies.

It is important to note that these \textit{ex vivo} results may not directly translate to \textit{in vivo} populations. A larger dataset would enable not only stronger differentiation between subpopulations, but also the use of statistical clustering methods (such as Gaussian mixture models) to determine subtype boundaries. Notably, the translation of DP to \textit{in vivo} can utilize ultrasound guidance which may improve the precision of the test, but may introduce a new set of challenges. 

In conclusion, DP may offer clinicians a minimally invasive, cost-effective, and precise tool that provides real-time information on thyroid nodule malignancy. Additionally, DP may be an applicable method to characterize disease in other tissues like breast or liver, ultimately reducing the number of avoidable surgeries and improving patient quality of life.

%%%%%%%%%%%%%%%% REFERENCES %%%%%%%%%%%%%%%

\clearpage % Clear all remaining figures and tables then start a new page

% The list of references goes after the main text and before the acknowledgements
% When preparing an initial submission, we recommend you use BibTeX, like this:
%
\bibliography{science_template} % for a file named science_template.bib
\bibliographystyle{sciencemag}

% After the paper has completed peer review and been revised ready for acceptance,
% you should comment out the lines above and copy-paste the contents of your .bbl
% file here instead. This will help ensure that our conversion software works correctly.
% Remember to re-run BibTeX first - check the timestamp!
%
% Example of the first three entries copy-pasted from science_template.bbl:
%
%\begin{thebibliography}{1}
%
%\bibitem{example}
%A.~N. {Author}, An example reference. \emph{Journal of Improbable Research}
%  \textbf{1}, 67 (2020).
%
%\bibitem{example2}
%F.~M. {Surname}, S.~{Author}, A second example. \emph{Interesting Research
%  Letters} \textbf{32}, 897 (2019).
%
%\bibitem{example_preprint}
%P.~{One}, P.~{Two}, P.~{Three}, {An unpublished preprint}. \emph{preprint}
%  (2021), arXiv:2101.12345.
%
%\end{thebibliography}

%%%%%%%%%%%%%%%% ACKNOWLEDGEMENTS %%%%%%%%%%%%%%%

\section*{Acknowledgments}
B.U. wishes to dedicate this paper to his sisters, Kristina and Alyssa, who were sources of inspiration for this work. The authors acknowledge the Newton-Wellesley Hospital Mass General Brigham (MGB) Pathology Team for their assistance with this project. J.B. would like to acknowledge the National Science Foundation (NSF) Graduate Research Fellowship Program as his primary funding source. H.V. acknowledges the support of the Department of Defense (DOD) through the National Defense Science \& Engineering Graduate (NDSEG) Fellowship Program. S.K.N. acknowledges the MIT Postdoctoral Fellowship Program for Engineering Excellence (PFPFEE). S.P. and E.K. would like to acknowledge the Ruane fund at Massachusetts General Hospital (MGH).
%\paragraph*{Author contributions:}
%\textcolor{red}{B.U. ...} 
\paragraph*{Competing interests:}
There are no competing interests to declare.
\paragraph*{Data, code and materials availability:}
All data and code are publicly available at the Cohen Mechanics Group GitHub (\textit{cohen-mechanics-group}).

%%%%%%%%%%%%%%%% END OF MAIN TEXT %%%%%%%%%%%%%%%

\newpage

%%%%%%%%%%%%%%%% START OF SUPPLEMENT %%%%%%%%%%%%%%%

% Figures, tables, equations and pages in the supplement are numbered S1, S2 etc.
\renewcommand{\thefigure}{S\arabic{figure}}
\renewcommand{\thetable}{S\arabic{table}}
\renewcommand{\theequation}{S\arabic{equation}}
\renewcommand{\thepage}{S\arabic{page}}
\setcounter{figure}{0}
\setcounter{table}{0}
\setcounter{equation}{0}
\setcounter{page}{1} % not 0 as \newpage already started a supplementary page
% References continue the numbering from the main text.

%%%%%%%%%%%%%%%% SUPPLEMENT TITLE PAGE %%%%%%%%%%%%%%%

\begin{center}
    \section*{Supplementary Materials for\\ \scititle}
\end{center}
    \author{
	% You can write out first names or use initials - either way is acceptable, but be consistent
	Brendan~M.~Unikewicz$^{1}$,
	Joseph~E.~Bonavia$^{1}$,
	Ekaterina~L.~Koelliker$^{2}$, \and
    Abigail~Klein$^{3}$,
	Katie~C.~Spaeth$^{1}$,
	Hannah~M.~Varner$^{1}$, \and
    S.~Kiana~Naghibzadeh$^{4}$,
    Yoav~Horev$^{5}$, 
	Sareh~Parangi$^{2\dagger}$, 
    Tal~Cohen$^{1,4\dagger}$\and
	% Additional lines of authors should be inserted using the \and command (not \\)
	% Institution list, in a slightly smaller font
    \newline \newline 
	\small$^{1}$Department of Mechanical Engineering, MIT, Cambridge, MA, USA.\and\ \newline 
	\small$^{2}$Department of Surgery, Massachusetts General Hospital and Harvard Medical School, Boston, MA, USA.\and\
    \small$^{3}$Department of Electrical Engineering and Computer Science, MIT, Cambridge, MA, USA.\and\ \newline 
    \small$^{4}$Department of Civil and Environmental Engineering, MIT, Cambridge, MA, USA.\and\ \newline
    \small$^{5}$Division of Otolaryngology, Head and Neck Surgery, University of Alberta, Edmonton, Alberta, Canada.\and\
	% Do not include street addresses, postal codes etc.
	%
	% Identify at least one corresponding author, with contact email address
    \newline

	\small$^\dagger$Corresponding author. Email: talco@mit.edu and sparangi@mgb.org\and
}    

\newpage

%%%%%%%%%%%%%%%% MATERIALS AND METHODS %%%%%%%%%%%%%%%

\section*{Materials and Methods}

\subsection*{Thyroid sample procurement and handling} In the course of this study, we performed a total of 36 DP tests in 10 different nodules. The number of tests in each nodule ranged between 2 to 7, depending on the size and location of the nodule, to ensure sufficient distance between insertion locations and from the nodule edge.

We followed a standard procedural protocol for thyroid sample and measurement. Excised human thyroids samples were procured at Newton-Wellesley Hospital in accordance with the IRB protocol 2008P001466. All methods were carried out in accordance with the IRB guidelines and regulations. %Thyroids that first underwent preoperative ultrasound and in some cases biopsy, and then required removal, underwent full thyroidectomies. 
Once the thyroidectomy was completed, the  thyroid specimen was transferred to the on-site pathology room where the MIT team was prepared to perform DP prior to  pathology evaluation. First, nodules were identified   in consultation by the clinical team which included a pathologist. Initial and final diagnoses, from the pre-operative ultrasound, nodule size characteristics, and the final pathology diagnoses, were kept in a secured database for reference. DP test were performed  immediately after surgical removal and testing occurred over the course of approximately one hour, depending on the number of nodules tested. Once testing was complete, the thyroid was then placed in formalin.

\subsection*{Experimental setup} Prior to a thyroid being received for testing, the experimental apparatus was prepared, on-site, at the Newton-Wellesley Hospital pathology room. The experimental apparatus was comprised of three distinct modules, each with their own Universal Serial Bus (USB) connections into a laptop computer: 1) the pressure sensing module, 2) the insertion module, and lastly, 3) the injection module (See Fig. \ref{fig:fig1-esi} in the Supplementary). 

\subsubsection*{1. Pressure sensing module} This module is comprised of a luer-tipped borosilicate glass 10$\mu$L syringe (Hamilton 10$\mu$L  Syringe) filled with Phosphate-buffered saline (PBS) at a 9:1 ratio with distilled water, directly connected to a luer-locking pressure sensor (PRESS-S-0000), capped with a standard 25 Gauge (25G) hypodermic needle that is 25.4mm in length. Error estimates for the pressure sensor, depending on the recorded value, had a specification of $\pm$2\% for pressures recorded within 0 to 41.4kPa (0 to 6psi), $\pm$3\% for 41.4 to 206.7kPa (6 to 30psi), and $\pm$3\% for 0 to -48.3kPa (0 to -7psi). Analog to Digital Conversion (ADC) from the pressure sensor was performed via a 24-Bit ADC (Adafruit NAU7802). The ADC was interrogated over Inter-integrated circuit (I2C) protocols via a micro-controller USB key (Adafruit Trinkey QT2040) connected via a Qwiic connection. The micro-controller USB key was then plugged into a laptop computer (Lenovo Thinkpad X1 Carbon) and supplied a nominal 5.00VDC. The micro-controller USB key, then regulates the 5.00VDC to 3.30VDC for the 24-Bit ADC. The 24-Bit ADC then regulates the 3.30VDC to 3.00VDC to power the pressure sensor. The pressure sensor's resolution was 0.2584mV/V/psi at a supply voltage of 3.00VDC leaving our effective system resolution at 0.7752mV/psi (0.1125mV/kPa). The micro-controller USB key was flashed with firmware to install the CircuitPython bootloader. The firmware file can be found at the \textit{cohen-mechanics-group} GitHub under, \textit{u2if\_trinkey\_qt2040\_rp2040.uf2}. Interfacing with the pressure sensing module (and insertion and injection modules) is the Visual Studio (VS) Code Integrated Development Environment (IDE). Specifically, within the VS Code IDE, Python 3.7.0 was used for all hardware communication and data acquisition across all modules.

\subsubsection*{2. Insertion module} This model is comprised of a linear stage actuator (RATTMMOTOR EBX1605), for raising the thyroid specimen into the needle during testing. Mounted to the linear stage actuator is the translation stage which was printed out of Polylactic acid (PLA) plastic filament (Bambu Lab PLA Basic). This linear stage assembly was directly powered by a stepper motor controller (Polulu Tic 36v4) and the stepper motor controller was powered via a 8.5A 24V AC/DC power supply connected to standard wall outlet ($\sim$110VAC). Communications with the stepper motor controller were handled over serial via a USB Type-A to Micro-B cable through the VS Code IDE with the same control software interfacing with the pressure sensing and injection modules. 

\subsubsection*{3. Injection module} This module is comprised of  a syringe microinjector (World Precision Instruments UMP3T) and the microinjector controller (World Precision Instruments MICROTOUCH 2T). Specifically, the syringe microinjector is connected to the microinjector controller via serial cable to the rear panel of the microinjector controller. Next, the microinjector controller is powered by a traditional wall outlet ($\sim$110VAC) and additionally connected to the host laptop computer via a USB A-to-B cable. On the microinjector controller, remote access is enabled, such that the host laptop computer can remotely access and program the microcontroller for controlled injections via the VS Code IDE through Python 3.7.0 for concurrent pressure measurement with the pressure sensing module. Once setup, the pressure sensing module, e.g. the glass syringe is loaded into the syringe microinjector and the experimental apparatus is setup. 

\subsection*{Thyroid nodule testing protocol}
The testing protocol strictly followed the five following steps:
\subsubsection*{1. DP device preparation and calibration test} Prior to receiving thyroid specimens, a water calibration test was performed. For this, with the glass syringe completely filled with  PBS solution, an  injection procedure was performed into a 1oz disposable plastic cup to measure the pressure drop along the needle. For this, the injection protocol was identical to the DP testing protocol with a liquid transfer rate  of 0.6$\mu$L/s up to a fluid volume of 9.2$\mu$L. By inspecting the obtained pressure-volume calibration curve for expected values we  confirmed there was no presence of air bubbles and that all components are operating as intended. Once the water calibration was completed, and the system was verified to have no air entrapment, the glass syringe was refilled with the PBS solution and thyroid tissue specimen testing could commence.

\subsubsection*{2. Thyroid assessment and preparation} Upon receiving a resected  thyroid, post-surgery, the thyroid was rested on a glass petri dish and placed on top of the linear stage assembly. Nodules were located with the assistance of the clinical team, and the number and choice of nodules tested were determined by the time available prior to tissue fixation.

\subsubsection*{3. Insertion procedure} With the device and thyroid ready for testing, the  computer was used to initialize the testing protocol, beginning with the insertion procedure. First, the linear stage assembly was translated upwards until the 25G hypodermic needle made first contact with the surface of the thyroid specimen, positioned directly over top the thyroid nodule.  Needle  insertion was then performed at a controlled insertion rate of 12.50mm/s, by elevating the translation stage  while concurrently  recording the pressure. Insertion stopped upon arriving at the desired depth (see insertion depths in Tables S1 and S2 of the Supplementary). Approximately one minute after the insertion was complete, the expansion procedure was performed. 

\subsubsection*{4. Expansion procedure} Following insertion, the injection protocol was performed. Specifically, fluid was transferred via the syringe microinjector at a rate of 0.6$\mu$L/s for a total injected fluid volume of 9.2$\mu$L.

\subsubsection*{5. Test completion and device sterilization} Upon completion of the DP test, the hypodermic needle  was removed from the thyroid nodule via translating the stage downwards. The hypodermic needle was then placed in a 10\% bleach solution, and the translation stage was moved up and down, while periodically withdrawing and expelling 10\% bleach solution ($\sim$0.5$\mu$L) to cleanse the outer and inner portions of the hypodermic needle. Afterwards, the glass syringe was fully purged of any remaining liquid. Once the glass syringe assembly was fully emptied into the PBS solution, the glass syringe assembly was refilled with PBS, a water calibration performed once again to verify all components were operating as intended without the introduction of air bubbles, and thyroid nodule testing could commence once again. To test a same thyroid nodule, the thyroid specimen would be repositioned and reoriented such that needle insertion sites into the thyroid nodule were unique.

\subsection*{Extraction of characteristic features from DP data}
 From the insertion and injection curves, four key features were extracted via MATLAB (see Fig. \ref{fig1} and Table S5). First, the cavity expansion curve was calibrated by subtracting the pressure drop along the needle. This calibration accounts for pressure losses between the sensor and the needle tip as fluid is expelled into the thyroid specimen. The peak insertion pressure, $c_1$, is defined as the maximum recorded pressure during insertion. The peak pressure during cavity expansion, denoted $c_2$, was determined using a prominence-based peak detection algorithm (MATLAB \textit{findpeaks} command). To distinguish physiologically meaningful rupture events from signal noise, only peaks with a prominence and height exceeding a threshold of $\frac{1}{20}$ of the mean pressure were retained. Among all detected peaks meeting these criteria, the peak with the largest amplitude was selected as $c_2$, and the corresponding fluid volume at that point was recorded as $c_3$. The total number of detected peaks satisfying the criteria, denoted $N$, was transformed as $c_4 = N + \sqrt{1 + N^2}$ to regularize the discrete count and accommodate true zeros. If no peaks met the detection criteria, $N$ was set to zero, and $c_2$ and $c_3$ were defined as the pressure and volume at 99\% of the total expansion duration, ensuring that end-of-test artifacts did not bias the measurement.

\subsection*{Principal Component Analysis (PCA)}
To perform PCA we first compute variance inflation factors (VIFs) to assess multicollinearity among $\log$ transformed input features (see Table S3 in Supplementary). All VIF values were below 5.00, ensuring acceptable levels of multicollinearity  \cite{kim2019multicollinearity} and supporting the stability of the covariance structure necessary for PCA.
We note that in the feature normalization procedure described in the main text,  we included an offset of $ 10^{-6}$ kPa to $c_1$ to ensure there was no zero-handling during the analysis of $c_1$. 

\subsection*{Logistic Regressions and Signed Euclidean Distances}
Logistic regression was employed as a probabilistic classifier using the log-transformed features projected into principal component space, with model parameters estimated via maximum likelihood. Distinct logistic regression models were fitted to address three tasks: (i) distinguishing benign from malignant samples using the first three principal components; (ii) a comparison between PTC and the combined OTC+FTC subtypes to assess global separability of PTC from other malignancies; and (iii) subtype-specific distinguishment comparing each malignant subtype pair (PTC vs. OTC, PTC vs. FTC, and OTC vs. FTC) to evaluate inter-subtype separability and significance. For each model, the full set of regression coefficients $\beta_0$ through $\beta_3$ are reported, enabling reconstruction of the three-dimensional decision plane defined by $\beta_0 + \beta_1 \text{PC}_1 + \beta_2 \text{PC}_2 + \beta_3 \text{PC}_3 = 0$. From each model, signed Euclidean distances were computed as the orthogonal projection onto the axis normal to the decision plane: $D = \frac{\beta_1 \text{PC}_1 + \beta_2 \text{PC}_2 + \beta_3 \text{PC}_3}{\sqrt{\beta_1^2 + \beta_2^2 + \beta_3^2}}.$

To quantify significance, two-sample $t$-tests were performed and to understand separability, the Hedges' $g$ was used to understand mean effect size. To provide readers with an intuitive visualization of subtype clustering independent of parametric assumptions, a k-Nearest Neighbors analysis ($k = 6$) was performed in the PC1-PC2 plane; this visualization is included for interpretive purposes only and was not used for statistical or separability analysis.

\clearpage 
\section*{Supplementary: Tables and Figures}

\begin{table}[!hb]
\centering
\footnotesize

%\begin{tabular}{
%    c
%    c
%    p{3.1cm}
%    p{4cm}
%    p{4.2cm}
%    c
%}
\caption{\textbf{Summary of pre-operative ultrasound and post-operative pathology of benign thyroid (T) nodules (N).} Approximate nodule size is represented as Anteroposterior (AP) $\times$ Transverse (TRV) $\times$ Sagittal (SG) dimensions. Additionally, total programmed injection depth ($\delta$) is shown. $\delta$ varies with the depth of the nodule within the thyroid specimen and with specimen repositioning and reorientation to avoid repeated measurements at the same site. Note, $\delta$ is reported for completeness and results within the paper do not depend on $\delta$.}
\label{tab:thyroid_pathology}
\begin{tabular}{
l
c
c
c
c
c
r
}
\textbf{\#} &
\textbf{T--N} &
\textbf{Pre-Operative Ultrasound} &
\textbf{Post-Operative Pathology} &
\textbf{AP $\times$ TRV $\times$ SG} (cm) &
\textbf{$\delta$} (mm)\\

1  & 1--1 & Bethesda II: Benign & thyroid follicular nodular disease &
1 $\times$ 0.9 $\times$ 0.9 to 4.1 $\times$ 3.9 $\times$ 3 & 10\\

2  & 1--1 & Bethesda II: Benign & thyroid follicular nodular disease &
1 $\times$ 0.9 $\times$ 0.9 to 4.1 $\times$ 3.9 $\times$ 3 & 10\\

3  & 1--2 & Bethesda II: Benign & thyroid follicular nodular disease &
1 $\times$ 0.9 $\times$ 0.9 to 4.1 $\times$ 3.9 $\times$ 3 & 10\\

4  & 1--2 & Bethesda II: Benign & thyroid follicular nodular disease &
1 $\times$ 0.9 $\times$ 0.9 to 4.1 $\times$ 3.9 $\times$ 3 & 15\\

5  & 1--3 & Bethesda II: Benign & thyroid follicular nodular disease &
1.7 $\times$ 1.1 $\times$ 0.9 & 5\\

6  & 1--3 & Bethesda II: Benign & thyroid follicular nodular disease &
1.7 $\times$ 1.1 $\times$ 0.9 & 9\\

7  & 1--3 & Bethesda II: Benign & thyroid follicular nodular disease &
1.7 $\times$ 1.1 $\times$ 0.9 & 8\\

8  & 2--1 & Bethesda III: Indeterminate & macrofollicular nodule &
3.2 $\times$ 3.2 $\times$ 2.7 & 15\\

9  & 2--1 & Bethesda III: Indeterminate & macrofollicular nodule &
3.2 $\times$ 3.2 $\times$ 2.7 & 20\\

10 & 2--2 & Biopsy not performed & diffuse follicular hyperplasia &
0.8 $\times$ 0.8 $\times$ 0.7 & 8\\

11 & 2--2 & Biopsy not performed & diffuse follicular hyperplasia &
0.8 $\times$ 0.8 $\times$ 0.7 & 8\\

12 & 2--2 & Biopsy not performed & diffuse follicular hyperplasia &
0.8 $\times$ 0.8 $\times$ 0.7 & 8\\

13 & 2--2 & Biopsy not performed & diffuse follicular hyperplasia &
0.8 $\times$ 0.8 $\times$ 0.7 & 8\\

14 & 2--2 & Biopsy not performed & diffuse follicular hyperplasia &
0.8 $\times$ 0.8 $\times$ 0.7 & 15\\
\end{tabular}

\end{table}

\vspace{-1em}

\begin{table}
\footnotesize
\caption{\textbf{Summary of pre-operative ultrasound results and post-operative pathology for malignant thyroid (T) nodules (N).} Approximate nodule size is represented as Anteroposterior (AP) $\times$ Transverse (TRV) $\times$ Sagittal (SG) dimensions, and including TNM staging and AJCC classification. Additionally, total programmed injection depth ($\delta$) is shown. $\delta$ varies with the depth of the nodule within the thyroid specimen and with specimen repositioning and reorientation to avoid repeated measurements at the same site. Note, $\delta$ is reported for completeness and results within the paper do not depend on $\delta$.}
\label{tab:thyroid_pathology_malignant}
\begin{tabular}{
lcccccc
}
\textbf{\#} &
\textbf{T--N} &
\textbf{Pre-Operative Ultrasound} &
\textbf{Post-Operative Pathology} &
\textbf{TNM \& AJCC} &
\textbf{AP $\times$ TRV $\times$ SG} (cm) &
\textbf{$\delta$} (mm)\\

1  & 3--1 & Bethesda III: Indeterminate & follicular thyroid carcinoma &
T1bNxMx; I &
1.6 $\times$ 1.4 $\times$ 0.9 & 5\\

2  & 3--1 & Bethesda III: Indeterminate & follicular thyroid carcinoma &
T1bNxMx; I &
1.6 $\times$ 1.4 $\times$ 0.9 & 5 \\

3  & 3--1 & Bethesda III: Indeterminate & follicular thyroid carcinoma &
T1bNxMx; I &
1.6 $\times$ 1.4 $\times$ 0.9 & 5\\

4  & 3--2 & Bethesda IV: Indeterminate & follicular thyroid carcinoma &
T1bNxMx; I & 
1.6 $\times$ 1.4 $\times$ 0.9 & 4\\

5  & 3--2 & Bethesda IV: Indeterminate & follicular thyroid carcinoma &
T1bNxMx; I &
1.6 $\times$ 1.4 $\times$ 0.9 & 4\\

6  & 4--1 & Bethesda VI: Malignant & papillary thyroid carcinoma &
T2N1aMx; II &
2.2 $\times$ 1.1 $\times$ 1 & 8\\

7  & 4--1 & Bethesda VI: Malignant & papillary thyroid carcinoma &
T2N1aMx; II &
2.2 $\times$ 1.1 $\times$ 1 & 6\\

8  & 4--1 & Bethesda VI: Malignant & papillary thyroid carcinoma &
T2N1aMx; II &
2.2 $\times$ 1.1 $\times$ 1 & 7\\

9  & 4--1 & Bethesda VI: Malignant & papillary thyroid carcinoma &
T2N1aMx; II &
2.2 $\times$ 1.1 $\times$ 1 & 8\\

10 & 4--1 & Bethesda VI: Malignant & papillary thyroid carcinoma &
T2N1aMx; II &
2.2 $\times$ 1.1 $\times$ 1 & 7\\

11 & 4--1 & Bethesda VI: Malignant & papillary thyroid carcinoma &
T2N1aMx; II &
2.2 $\times$ 1.1 $\times$ 1 & 8 \\

12 & 4--1 & Bethesda VI: Malignant & papillary thyroid carcinoma &
T2N1aMx; II &
2.2 $\times$ 1.1 $\times$ 1 & 5\\

13 & 5--1 & Bethesda VI: Malignant & papillary thyroid carcinoma &
T1bN1aMx; II &
1.1 $\times$ 1.0 $\times$ 0.7 & 6\\

14 & 5--1 & Bethesda VI: Malignant & papillary thyroid carcinoma &
T1bN1aMx; II &
1.1 $\times$ 1.0 $\times$ 0.7 & 5 \\

15 & 5--1 & Bethesda VI: Malignant & papillary thyroid carcinoma &
T1bN1aMx; II &
1.1 $\times$ 1.0 $\times$ 0.7 & 5 \\

16 & 5--1 & Bethesda VI: Malignant & papillary thyroid carcinoma &
T1bN1aMx; II &
1.1 $\times$ 1.0 $\times$ 0.7 & 20 \\

17 & 6--1 & Bethesda IV: Indeterminate & oncocytic thyroid carcinoma &
T2NxMx; II &
3.7 $\times$ 3.2 $\times$ 2.6 & 20 \\

18 & 6--1 & Bethesda IV: Indeterminate & oncocytic thyroid carcinoma &
T2NxMx; II &
3.7 $\times$ 3.2 $\times$ 2.6 & 15 \\

19 & 6--1 & Bethesda IV: Indeterminate & oncocytic thyroid carcinoma &
T2NxMx; II &
3.7 $\times$ 3.2 $\times$ 2.6 & 13 \\

20 & 6--1 & Bethesda IV: Indeterminate & oncocytic thyroid carcinoma &
T2NxMx; II &
3.7 $\times$ 3.2 $\times$ 2.6 & 14 \\

21 & 6--1 & Bethesda IV: Indeterminate & oncocytic thyroid carcinoma &
T2NxMx; II &
3.7 $\times$ 3.2 $\times$ 2.6 & 13 \\

22 & 6--1 & Bethesda IV: Indeterminate & oncocytic thyroid carcinoma &
T2NxMx; II &
3.7 $\times$ 3.2 $\times$ 2.6 & 14 \\
\end{tabular}
\end{table}

\newpage

\begin{table}
\centering
\footnotesize
\caption{\textbf{Multicollinearity evaluation of the features identified.} Multicollinearity assessment (VIFs) of log-transformed features values $C_i$ and PCA loadings for PC1-3 reconstruction and signed Euclidean distance from the plane in log-space.}
\label{tab:pca_planes}
\begin{tabular}{lccccc}
\textbf{Feature} & \textbf{VIF} & \textbf{PC1 Loading} & \textbf{PC2 Loading} & \textbf{PC3 Loading} & \textbf{D}\\
$C_1$ & 1.25 & 0.4377 & 0.6262 & 0.4297 & 0.4359\\
$C_2$ & 1.21 & -0.3496 & 0.7610 & -0.1719 & 0.5552\\
$C_3$ & 1.27 & -0.5597 & -0.1182 & 0.8175 & 0.7069\\
$C_4$ & 1.45 & 0.6107 & -0.1215 & 0.3428 & 0.0437\\
\end{tabular}

\end{table}

\clearpage
\begin{table} 
\centering
\small
\setlength{\tabcolsep}{4pt}
\caption{\textbf{Measures of statistical significance as well as values for 3D plane reconstruction.} Signed distance $p$- and $g$-values for benign versus malignant nodules as well as logistic regression coefficients for the 3D decision boundary.}
\label{tab:pairwise_comparison}
\begin{tabular}{lccccccr}
\textbf{Comparison} & \textbf{$p$-value} & \textbf{Hedges' $g$} & \multicolumn{4}{c}{\textbf{Logistic regression $\beta$ coefficients}} \\

- & - & - & $\beta_0$ & $\beta_1$ (PC1) & $\beta_2$ (PC2) & $\beta_3$ (PC3) \\
Benign vs Malignant & 0.000001 & 1.94 & 0.8743 & 0.8604 & 1.846 & -1.179 \\
Nodule Averaged: Benign vs Malignant & 0.0009 & 2.90 & --- & --- & --- & --- \\
\end{tabular}
\end{table}

\clearpage
\begin{table}
\centering
\caption{\textbf{Log-transformed features values $C_i$ for benign nodule data.} Final pathology results shown against the log-transformed feature values ascertained through Digital Palpation.}
\label{tab:log-transformed-features-benign}
\begin{tabular}{
lccccc
}
\textbf{\#} &
\textbf{Final Pathology} &
\textbf{$C_1$} &
\textbf{$C_2$} &
\textbf{$C_3$} &
\textbf{$C_4$} \\
1   & Benign & -0.1448 & 1.4213 & 2.2755 & 3.5272 \\
2   & Benign & -0.5521 & 1.4611 & 2.2948 & 1.8184 \\
3   & Benign & 0.1215 & 1.0580 & 2.7559 & 1.8184 \\
4   & Benign & 0.4042 & 1.8899 & 3.0434 & 0.0000 \\
5   & Benign & -2.2694 & 0.8374 & 2.1281 & 3.3335 \\
6   & Benign & -0.8831 & 3.0272 & 3.0433 & 0.0000 \\
7   & Benign & -0.9384 & 2.8175 & 3.0434 & 0.0000 \\
8   & Benign & 0.6554 & 2.1347 & 3.0183 & 1.8184 \\
9   & Benign & 0.6983 & 1.7797 & 2.4879 & 3.5272 \\
10  & Benign & -1.5238 & 1.1453 & 1.7960 & 2.8934 \\
11  & Benign & 0.8879 & 1.1204 & 1.4794 & 4.2487 \\
12  & Benign & -2.0370 & 2.3880 & 2.8350 & 1.8184 \\
13  & Benign & 2.0395 & 2.8689 & 1.8154 & 1.4436 \\
14  & Benign & 1.5092 & -0.5498 & 2.9267 & 4.4069 \\
\end{tabular}
\end{table}

\begin{table}
\centering
\caption{\textbf{Log-transformed features values $C_i$ for malignant nodule data.} Final pathology results shown against the log-transformed feature values ascertained through Digital Palpation.}
\label{tab:log-transformed-features-malignant}
\begin{tabular}{
lccccc
}
\textbf{\#} &
\textbf{Final Pathology} &
\textbf{$C_1$} &
\textbf{$C_2$} &
\textbf{$C_3$} &
\textbf{$C_4$} \\
1  & Malignant & 0.3178 & 2.8640 & 2.1536 & 0.8814 \\
2  & Malignant & -0.1690 & 2.9954 & 2.1585 & 1.8184 \\
3  & Malignant & -2.8602 & 4.0063 & 2.4407 & 1.4436 \\
4  & Malignant & -0.0402 & 2.1184 & 1.9039 & 0.8814 \\
5  & Malignant & -0.3390 & 1.4154 & 1.7628 & 1.8184 \\
6  & Malignant & 2.1707 & 1.2167 & 2.8771 & 4.3309 \\
7  & Malignant & 0.9138 & 2.6220 & 2.0531 & 2.0947 \\
8  & Malignant & 2.1765 & 2.2847 & 1.5309 & 1.8184 \\
9  & Malignant & 2.9618 & 1.6357 & 1.7091 & 2.0947 \\
10  & Malignant & 3.0949 & 3.0242 & 1.6238 & 2.4918 \\
11  & Malignant & 1.7636 & 1.2407 & 1.4418 & 3.9124 \\
12  & Malignant & 3.1249 & 5.2664 & 2.9242 & 2.4918 \\
13  & Malignant & 1.4476 & 3.1259 & 1.5768 & 4.5851 \\
14  & Malignant & 1.4431 & 3.6770 & 1.7439 & 4.0257 \\
15  & Malignant & 2.0510 & 4.1512 & 1.8213 & 2.6441 \\
16  & Malignant & 1.3855 & 0.9536 & 1.4383 & 2.7765 \\
17  & Malignant & 1.2895 & 0.3932 & 1.4829 & 3.1798 \\
18  & Malignant & 1.2145 & 1.5858 & 1.6051 & 3.2596 \\
19  & Malignant & 1.1911 & 1.4349 & 1.6728 & 2.0947 \\
20  & Malignant & 1.3552 & 1.5627 & 1.5746 & 2.6441 \\
21  & Malignant & 0.8540 & 0.7222 & 1.4378 & 3.4667 \\
22  & Malignant & 0.7333 & 1.4053 & 1.4484 & 3.4667 \\
\end{tabular}
\end{table}

\begin{figure}
    \centering
    \includegraphics[width=0.5\textwidth]{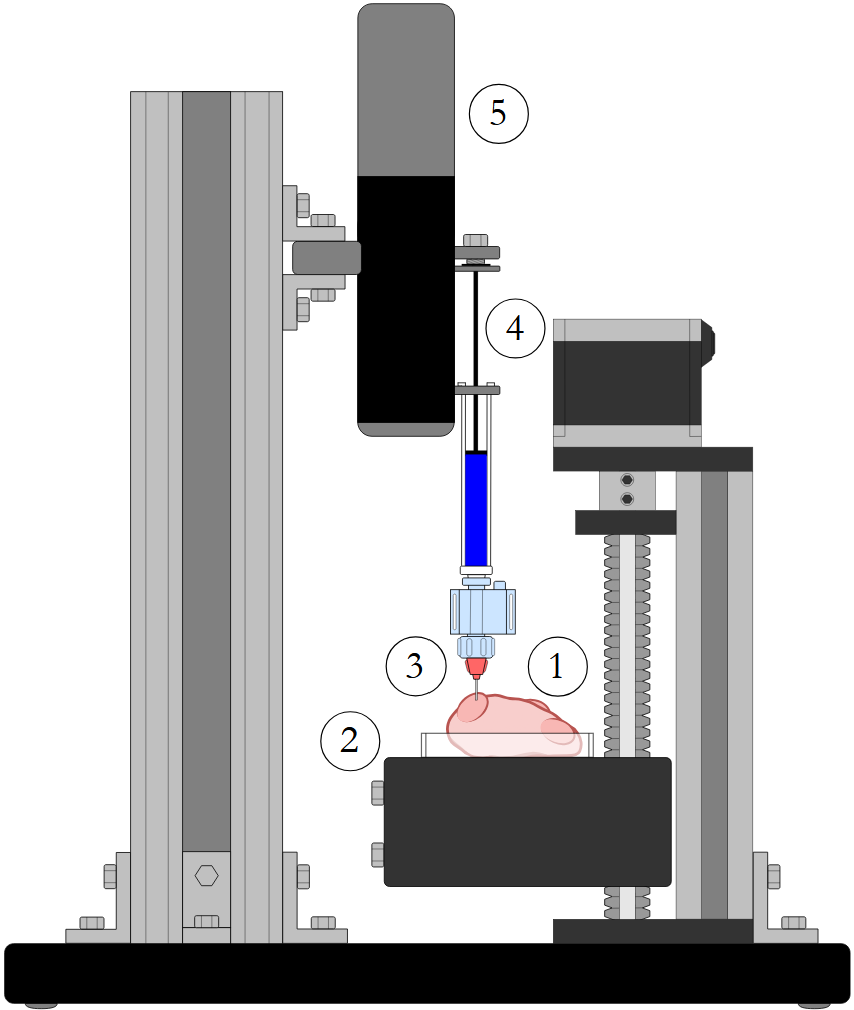}
    \caption{\textbf{Digital Palpation experimental setup.} A thyroid specimen (1) containing nodules suspected of malignancy is positioned atop a vertically actuated translation stage (2) immediately after being resected from surgery. The stage raises the thyroid (1) into a fixed beveled needle (3), which penetrates the nodule. Following insertion, the cavity expansion protocol is initiated by displacing a syringe plunger (4) downward, injecting fluid at a controlled rate, via a high precision microinjector (5), into the tissue to form a fluid cavity.}
    \label{fig:fig1-esi}
\end{figure}

\begin{figure}[!h]
    \centering
    \includegraphics[width=0.75\textwidth]{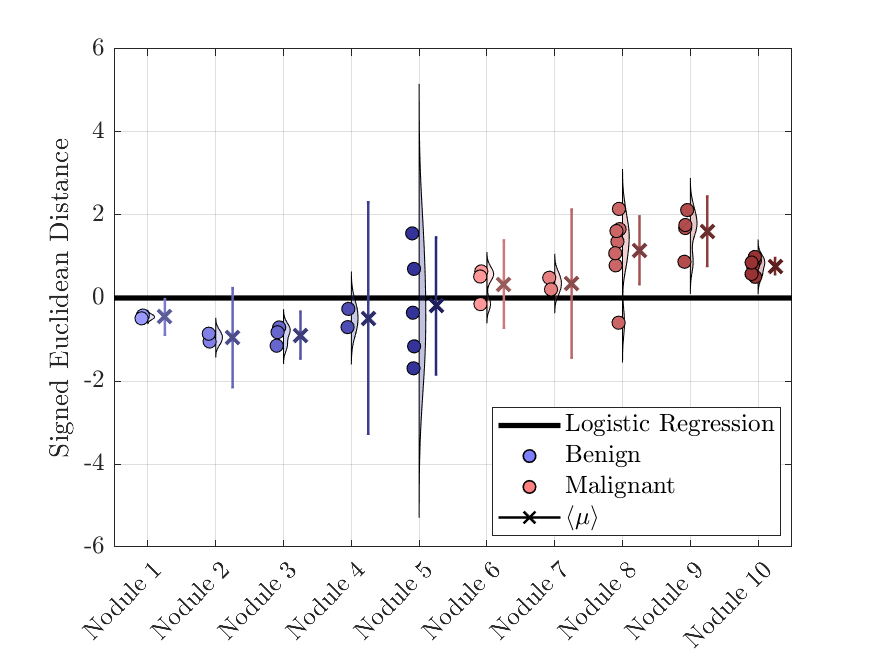}
    \caption{\textbf{Per nodule data collected during Digital Palpation.} Signed euclidean distances for each nodule, corresponding with Tables S1 and S2, with average values per nodule (and their equivalent 95\% confidence limits) shown.}
    \label{fig:fig2-esi}
\end{figure}

%%%%%%%%%%% CAPTIONS FOR OTHER SUPPLEMENTARY FILES %%%%%%%%%%

%%%%%%%%%%%%%%%% SUPPLEMENTARY REFERENCES %%%%%%%%%%%%%%%

% Do NOT include a reference list in the supplement.
% All references must be in a single list at the end of the main text.
% The copyeditors will ensure that the correct reference list appears with each version of the paper
% (print, HTML, PDF, mobile app, metadata for bibliographic databases etc.)

\end{document}